\documentclass[twocolumn,aps,pre,superscriptaddress]{revtex4}
\usepackage{hyperref}
\usepackage{amsmath, amssymb, graphicx, xcolor}
\usepackage{bm}

\begin{document}

\title{Binomial maps: stochastically evolving iterated integer maps for finite populations}


\author{Snehal M. Shekatkar}
\email[]{snehal@inferred.in}
\affiliation{School of Computing and Data Sciences, FLAME University, Pune 412115, India}



\begin{abstract}
Many models of population dynamics are formulated as deterministic iterated maps although real populations are stochastic. This is justifiable in the limit of large population sizes, as the stochastic fluctuations are negligible then. However, this also makes it challenging to use the same models for small populations where finite size effects like demographic noise and extinction cannot be ignored. Moreover, adding noise to the equations does not solve this problem as it can only represent the environmental stochasticity. An approach, sometimes used in ecological literature, but surprisingly uncommon in dynamical systems community, is \emph{Binomial maps}, which allow stochastic evolution of deterministic iterated map models of population. Here we present their formulation in a way so as to make their connection to the agent-based models explicit, and demonstrate it for the Logistic and Ricker maps. We also show that the Binomial maps are not completely equivalent to their deterministic counterparts, and derive sufficient conditions under which the equivalence holds. This approach enables rigorous finite-population analysis within familiar map-based models, bridging the deterministic map models and stochastic agent-based models.
\end{abstract}

\maketitle



Many models in population dynamics are constructed as deterministic discrete-time dynamical systems or \emph{iterated maps} \cite{Kot2001Elements}. Some examples include the famous Logistic map \cite{May1976}, Ricker model \cite{Ricker1954}, Nicholson-Bailey model \cite{NicholsonBailey1935}, Baranov catch equation \cite{Quinn2003BaranovReview}, and Beverton-Holt model \cite{Hassell1975BevertonHolt}. Although the dynamics of the individuals constituting the population of interest is always stochastic, these models are formulated to be deterministic. The reason for this is that when the underlying population size is practically infinite, the sizes of fluctuations relative to the mean are negligible in this continuum limit. 

While these models are simple to study and simulate, they fail to be applicable when the underlying population size is small, as then the relative size of the fluctuations cannot be ignored. Taking into account fluctuations is important not only because they lead to finite size outcomes like extinction, but also because they represent the inherent demographic noise which tells us how much variability to expect in the population size at any given time \cite{Ovaskainen2010ExtinctionReview, Lande1993ExtinctionRisk, MelbourneHastings2008Nature, Doering2005ExtinctionTimes, Engen1998Concepts, Engen2001TimeToExtinction, Dennis1991Estimation}.

\emph{Agent-based models} is an approach that can work in both small and large population limits and does not suffer from the problems suffered by the deterministic map models. However, apart from being computationally inefficient, it requires the knowledge of stochastic microscopic laws of motion for the individuals or agents in the population. Since these laws are not explicitly present in the equations of the deterministic map models, they cannot be coverted to agent-based models to simulate finite populations \cite{Ovaskainen2010ExtinctionReview, GrimmRailsback2005IBM, DeAngelisMooij2005Review}. 

For the continous-time models, this problem can be easily addressed by either stochastically integrating deterministic equations either using time-driven Monte-Carlo \cite{parker_mean_2010} or the faster event-driven Gillespie algorithm \cite{gillespie1977exact}. These approaches are computationally efficient when we are interested only in the macroscopic behavior of the system and not in the motions of individual agents, but they require knowledge of the underlying processes. 

For discrete-time systems, an analogous approach called \emph{Binomial maps} exists, and is sometimes used in ecological and epidemiological modeling \cite{melbourne_extinction_2008, lefevre_chain_2021}. Unfortunately, in spite of their strength in handling finite populations, they have not become a common tool in the dynamical systems studies. 
Here we not only describe their structure, but also put their formulation on a firm theoretical footing by making their connection to the agent-based models explicit. 
We also show that, somewhat surprisingly, a deterministic integer map and its corresponding Binomial map are not exactly equivalent in the sense that the average of the stochastic trajectories is in general not the same as the deterministic trajectory of the original map. Nevertheless, as we prove, under appropriate conditions, the equivalence does hold for most of the commonly studied iterated maps. For these systems, which includes the Logistic map, the Binomial maps interpolate between discrete-time agent-based models and deterministic maps. 

\emph{Formulating Binomial maps}. Consider the deterministic iterated map, ${\mathbf X}(t+1) = \mathbf{f}(\mathbf{X}(t), \Theta)$ where the vector $\mathbf{X}(t)$ denotes the \emph{average} state at time $t$ and $\Theta$ denotes the parameters of the system. Since each component $X_i$ of ${\mathbf X}$ represents an average population, it is always nonnegative, but can take continuous values. Also, in many models, there is a maximum value that it cannot exceed at any given time, which depends on parameters $\Theta$. For example, in many ecological models, the population cannot exceed the carrying capacity. 
In some systems, this maximum value may even depend on the current value of ${\mathbf X}$, and hence could be time-dependent. For example, in the discrete-time SIR model, the largest number of new infections at any time cannot exceed the number of susceptibles in the previous time step which itself depends on time. Let us thus define $\mathbf{A}_t=\mathbf{\phi}(\mathbf{X}(t), \Theta)$ to be the \emph{Availability vector}, whose components represent these maximum possible values at time $t$. If there is no finite bound for a certain component $X_{i}$ then we set $A_i\to\infty$. 

Since we are assuming that this system represents the average behavior of a system of stochastic agents, the actual state of the system is given by a random vector ${\mathbf X}_b(t)$ whose components are non-negative integers. In the model ${\mathbf X}(t+1) = \mathbf{f}(\mathbf{X}(t), \Theta)$, the state vector at time $t$ determines ${\mathbf X}(t+1)$, the state vector at time $t+1$. The core assumption of Binomial maps formulation is that instead of determining $\mathbf{X}_{b}(t+1)$ directly, the state $\mathbf{X}_{b}(t)$ instead determines only its \emph{mean} or the \emph{expected value} $\mathbb{E}[\mathbf{X}_b(t+1)]$. The actual value of each component ${X}_{b,i}(t+1)$ is then drawn from a probability distribution with mean $\mathbb{E}[X_{b,i}(t+1)]$. To complete the formulation, we must now decide the form of this probability distribution. To do this, observe that we are assuming that $\mathbf{X}_b$ is representing a system of agents. Since in the deterministic map model the agents whose population is represented by the same component of $\mathbf{X}$ follow the same rule of evolution, these agents must be identical. Thus the probability that out of maximum $A_{i,t}$ agents, when there are $\mathbb{E}[X_{b,i}(t+1)]$ ``successes'' on average, each of the $A_{i,t}$ available agents is a Bernoulli random variable with success probability $\mathbb{E}[X_{b,i}(t+1)]/A_{i,t}$. This means that the total number of successes is a Binomial random variable $X_{b, i}(t+1) \sim \text{Binomial}(A_{i,t}, \mathbb{E}[X_{b,i}(t+1)/A_{i,t}]$. Thus, the evolution equation for the $i$'th component of the corresponding Binomial map is given by:
\begin{equation}
    \label{eq:stochastic_discrete}
    \begin{aligned}
    &\mathbb{E}[X_{b,i}(t+1)] = f_i({\mathbf X}_b(t), {\mathbf A}_t)\\
    &X_{b,i}(t+1) \sim \text{Binomial}\left(A_{i,t}, \frac{\mathbb{E}[X_{b,i}(t+1)]}{A_{i,t}}\right)
    \end{aligned}
\end{equation}
The functional form of ${\mathbf A}_{t} = {\mathbf \phi}(X_t^{(B)}, K)$ would depend on the specifics of the system, but can be easily determined as we demonstrate later in the paper. 
The size of the fluctuations in $X_{b,i}(t+1)$, as measured by the standard deviation of the Binomial distribution in this equation, is:
\begin{equation}
    \begin{aligned}
    \sigma = \sqrt{\mathbb{E}[X_{b,i}(t+1)](1-\mathbb{E}[X_{b,i}(t+1)]/A_{i,t})}
    \end{aligned}
    \label{eq:}
\end{equation}
 Or the size of the \emph{fluctuations relative to the mean} is: 
\begin{equation}
\sigma_{rel} = \frac{\sigma}{\mathbb{E}[X_{b,i}(t+1)]} = \sqrt{\frac{(1-\mathbb{E}[X_{b,i}(t+1)]/A_{i,t})}{\mathbb{E}[X_{b,i}(t+1)]}}
\end{equation}
Thus, unlike the models in which noise is added externally, the size of the fluctuations is not fixed and depends on the average size of the population as well as the availability at time $t$. In particular, $\sigma_{rel}$ is close to zero when $\mathbb{E}[X_{b,i}(t+1)]$ is large, which can happen only if $A_{i,t}$ is large as $\mathbb{E}[X_{b,i}(t+1)]/A_{i,t} \leq 1$. But $A_{i,t}$ can be large only when the number of agents is large, and hence the fluctuations can go to zero in the limit of large population size resulting in a smooth trajectory in the phase space. Note however that when $A_{i,t}$ is large but $\mathbb{E}[X_{b,i}(t+1)]$ is small, $\sigma_{rel}$ can be large. This is what is desired from the realistic modelling of changes in population since demographic fluctuations must dominate the underlying deterministic dynamics when the population is small but they should go to zero when the population size is large.

\begin{figure}[ht]
    \begin{center}
        \includegraphics[width=0.95\columnwidth]{"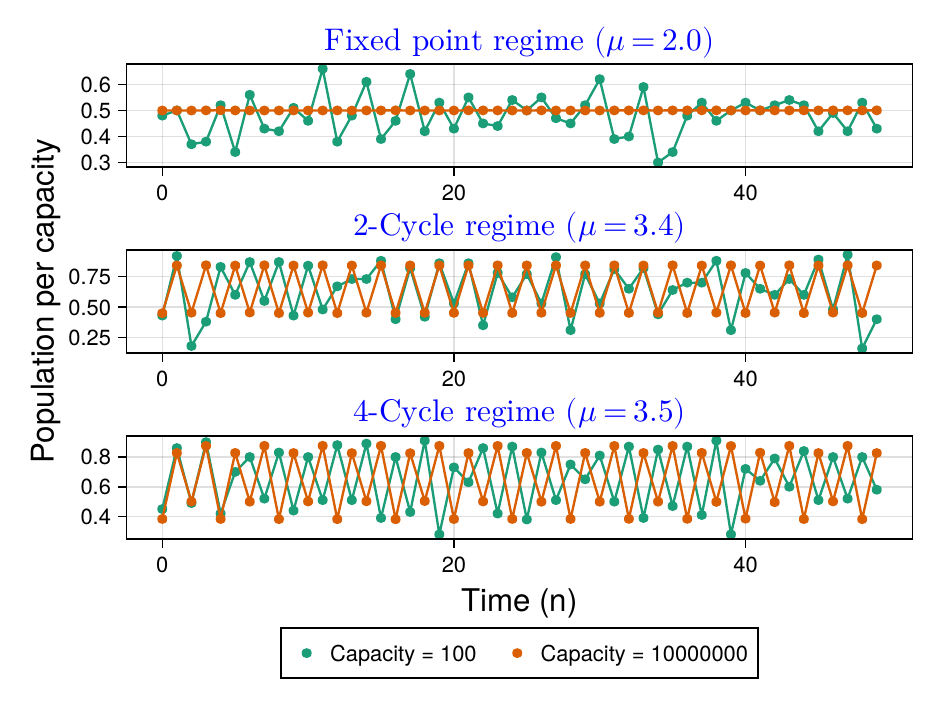"}
        \caption{\label{ts3r} Time series of the Binomial Logistic map for two different capacities in three different dynamical regimes. Fluctuations in the population present for small capacity can be seen to die down as capacity becomes large, and the map approaches the standard logistic map.}
    \end{center}
\end{figure}

\begin{figure}[ht]
       \includegraphics[width=0.95\columnwidth]{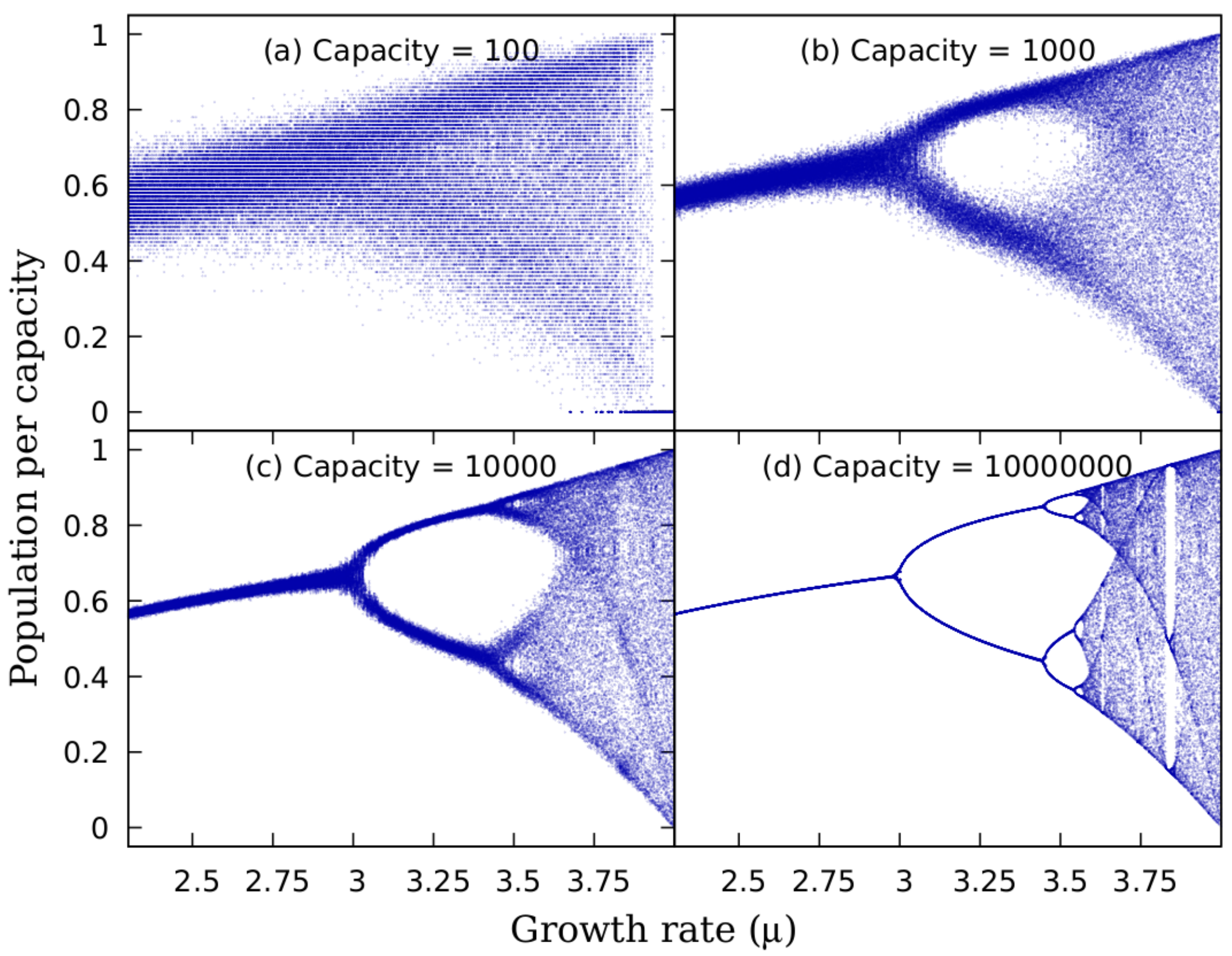}
       \caption{\label{bifur} The bifurcation diagram of the Binomial Logistic map for four different capacities. As can be clearly seen, the bifurcation diagram approaches the bifurcation diagram of the standard Logistic map as the capacity increases.}
\end{figure}

\emph{Binomial Logistic map}. As a concrete example, consider the Logistic map, a model of the population dynamics in an environment with limited resources so that maximum $K$ individuals can exist at any given time:
\begin{equation}
    X_{t+1} = \mu X_t\left(1-\frac{X_t}{K}\right)
\end{equation}
Thus, $A_t = K$ for all $t \geq 0$, and Eq(\ref{eq:stochastic_discrete}) reduces to:
\begin{equation}
    \label{eq:stochastic_logistic}
    \begin{aligned}
        &\mathbb{E}[X_{t+1}] = \mu X_b(t)\left(1-\frac{X_b(t)}{K}\right)\\
        &X_b(t+1) \sim \text{Binomial}\left(K, \frac{\mathbb{E}[X_{b}(t+1)]}{K}\right)
    \end{aligned}
\end{equation}
Later we will analytically prove that the dynamics of the stochastic map \ref{eq:stochastic_logistic} when expressed in terms of $x=X_b/K$ becomes indistinguishable from the deterministic map when the capacity $K$ is large. For now we can see this numerically by looking at the time series and the bifurcation diagram as shown in Fig.~\ref{ts3r} and Fig.~\ref{bifur}.
We also note that, because the Binomial random variable is allowed to take the zero value, it is possible at any time step for the population to go extinct without any external factor. The likelihood of this is high when the mean of the distribution $\mathbb{E}[X_{b}(t+1)]$ is small which is more likely to be small when the capacity $K$ itself is small. This means that a population with low capacity $K$ has higher probability of going extinct. This is consistent with the behavior of real populations because if the number of individuals is small and most individuals fail to reproduce by chance or if their offsprings fail to survive, the population can go extinct without any external intervention. 

\begin{figure}
    \includegraphics[width=0.8\columnwidth]{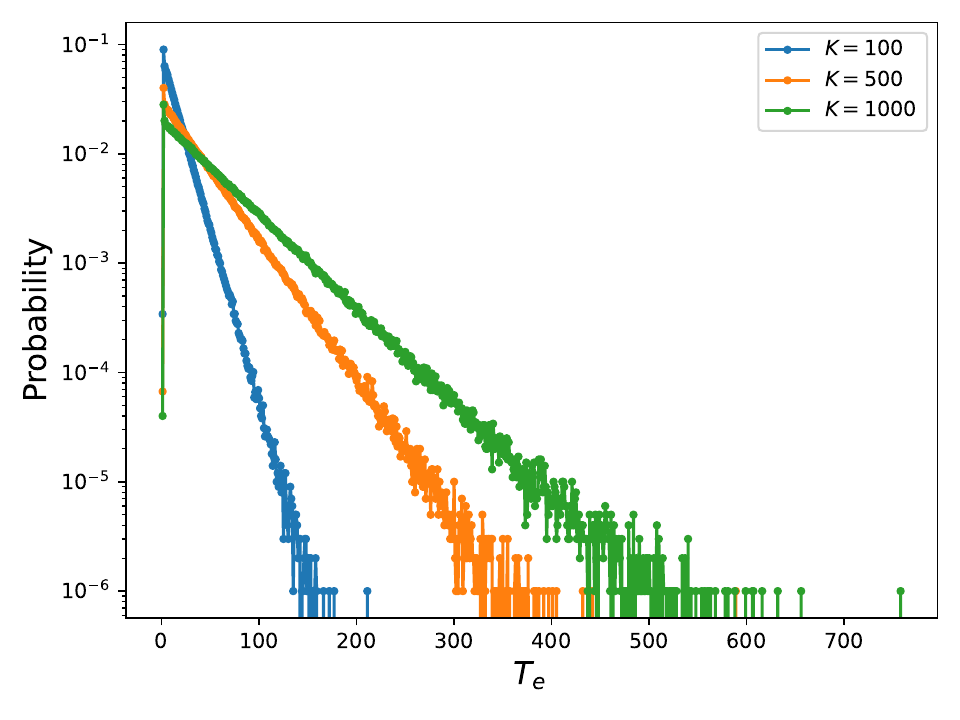}
    \caption{\label{logistic_extinction} Distributions of extinction times for the Binomial Logistic map for different carrying capacities $K$ with $\mu=4$ which corresponds to the chaotic regime in the original map. The distributions can be seen to be discrete-exponential (notice the log-scale on the vertical axis) except for $T_e=1$ and $T_e=2$. Each value is obtained by starting with an inital population size chosen uniformly randomly from the set $\{1, 2, \dots, K\}$, and then iterating the map until population reaches zero.}
\end{figure}

\emph{Extinction times for finite populations}. One of the important applications of the Gillespie algorithm is the estimation the extinction times for finite populations \cite{parker_mean_2010, dobrinevski_extinction_2012, el_chaal_stochastic_2023, korngut_weighted-ensemble_2025, thakur_controlling_2023}. This is done by generating a large number of stochastic trajectories and the extinction time is recorded for each trajectory. For a sufficiently large sample of trajectories this gives the complete distribution of the extinction time from which mean extinction time can be computed. The exactly same strategy can be used to estimate the extinction times for Binomial maps. 

Fig.~\ref{logistic_extinction} shows the application of this method to compute the probability distribution of the extinction time for the Binomial Logistic map with $\mu=4$. This practical utility of the formalism adds a new dimension to its value, highlighting it not just as a conceptual fix, but as a tool for extracting meaningful statistical insights about dynamical systems.

\begin{figure}
    \begin{center}
    \includegraphics[width=0.8\columnwidth]{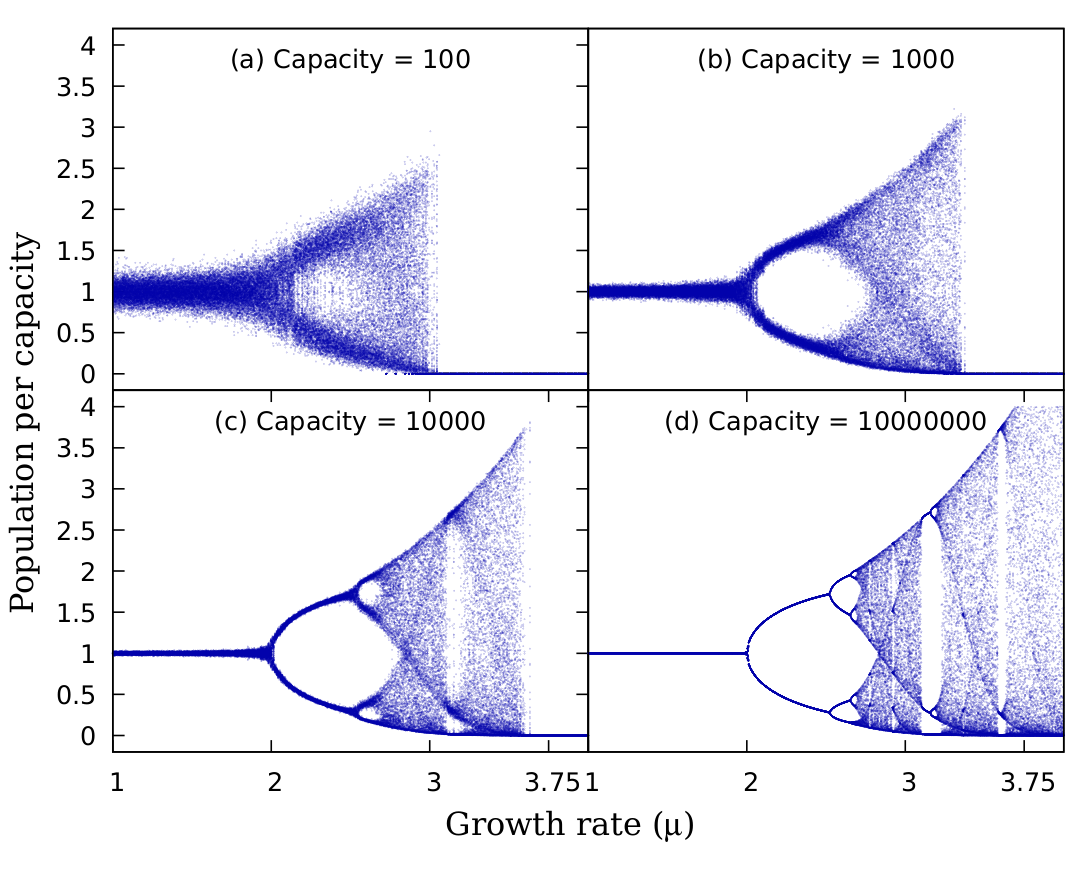}
    \caption{\label{fig:ricker_bifur} The bifurcation diagram for the Binomial Ricker map for different carrying capacities. As explained in the text, the lack of finite upper bound can be incorporated in the corresponding Binomial map by using the Poisson distribution instead of the binomial distribution. Also, although in the limit of large population the map converges to the original deterministic map, it still accommodates extinction.}
    \end{center}
\end{figure}
\emph{Binomial Ricker Map}.
As a second example, we consider the Ricker model whose evolution equation is:
\begin{equation}
    \label{eq:ricker}
    \begin{aligned}
        X_{t+1} = X_t \exp\left[{\mu\left(1-\frac{X_t}{C}\right)}\right]
    \end{aligned}
\end{equation}
In this model, there is no upper limit to the number of individuals although there is a parameter $C$, also called carrying capacity, and whenever population goes above it, it is exponentially quenched to a lower value. But since population is allowed to exceed it, this is an example of a model with infinite availability $A_t=\infty$. Thus, to construct the corresponding Binomial map, we need to modify Eq(\ref{eq:stochastic_discrete}) by taking limit $A_t\to\infty$. It is easy to see that in this case the distribution $\text{Binomial}(A_t, \mathbb{E}[X_b]/A_t)$ converges to the Poisson distribution with mean $\mathbb{E}[X_b]$. Hence, when $A_t\to\infty$, the corresponding Binomial map is given by:
\begin{equation}
    \label{eq:stochastic_discrete_poisson}
    \begin{aligned}
    &\mathbb{E}[X_{b}(t+1)] = f(X_b(t), C)\\
    &X_{b}(t+1) \sim \text{Poisson}\left(\mathbb{E}[X_b(t+1)]\right)
    \end{aligned}
\end{equation}
Hence, the Binomial map corresponding to the Ricker model (\ref{eq:ricker}) is given by:
\begin{equation}
    \label{eq:ricker_poisson}
    \begin{aligned}
        &\mathbb{E}[X_b(t+1)] = X_b(t) \exp\left[{\mu\left(1-\frac{X_b(t)}{C}\right)}\right]\\
        &X_b(t+1) \sim \text{Poisson}\left(\mathbb{E}[X_b(t+1)]\right)
    \end{aligned}
\end{equation}
For this reason, the maps with infinite availability are also called \emph{Poisson maps}.
The bifurcation diagram of Binomial Ricker map for various capacities is shown in Fig.~\ref{fig:ricker_bifur}, and similar to Binomial Logistic map, relative fluctuations can be seen to go to zero in the limit of infinite capacity.

\emph{Non-equivalence of Binomial maps with the deterministic maps}. 
Although the Binomial maps are constructed just by replacing the deterministic state variable ${\mathbf X}$ of a deterministic integer map by a stochastic state variable ${\mathbf X}_b$, it turns out that they are not completely equivalent as we explain now. This non-equivalence has been mentioned in the context of stochastic evolution of continuous-time systems in \cite{gustafsson_when_2013}. Here we formally prove it for the discrete-time systems using the Law of total expectation, and moreover, derive conditions under which the equivalence holds. Suppose we construct a Binomial map corresponding to a given deterministic integer map, and then evolve both of them starting with the same initial condition. For the deterministic map, we get a single trajectory, and for the Binomial map, there is an ensemble of trajectories with that initial condition. It is natural to expect that the \emph{average} of the Binomial map trajectories is same as the deterministic map trajectory in the limit of infinite number of agents.
Unfortunately this is not in general true for a subtle reason related to noncommutativity of the expectation value operator and a nonlinear function. At the same time, it is true that for a large class of commonly studied systems, including the ones we saw above, the equivalence holds in the limit of infinite availability. 

Suppose we iterate the 1d deterministic map $X(t+1)=f(X(t), \Theta)$ and the corresponding Binomial map starting from the same initial condition $X_0$ at $t=0$. For the deterministic map, the value at $t=1$ is $X(1) = f(X_0)$ whereas for the Binomial map, $X_b(1)$ takes one of the $A_0+1$ values from the set $\{0, 1, 2, \dots, A_0\}$ such that $\mathbb{E}[X_b(1)]=f(X_0)=X(1)$. Hence, at time $t=1$, the average of the Binomial map trajectories is same as the value of the deterministic map variable. At $t=2$, the deterministic map outputs $X(2) = f(X(1)) = f(\mathbb{E}(X_b(1)))$, but for the Binomial map, $\mathbb{E}[X_b(2)] = \mathbb{E}[\mathbb{E}[X_b(2)|X_b(1)]] = \mathbb{E}[f(X_b(1))]$. Since in general $f(\mathbb{E}[X]) \neq \mathbb{E}[f(X)]$, the average of the stochastic trajectories of Binomial map is not the same as the trajectory generated by the deterministic map.  

However, as we demonstrate below, for many systems the gap $\delta J = f(\mathbb{E}[X_b(t)])-\mathbb{E}[f(X_b(t))]$, called the Jensen gap, goes to zero in the limit $A_t\to \infty$, and hence for these systems, Binomial maps interpolate between the determinstic maps and the agent-based models. To derive a sufficient for this, let us assume that $f$ is analytic everywhere, i.e. its Taylor series expansion exists around any point on the real-line. We can then write: 
\begin{equation}
    \begin{aligned}
    \mathbb{E}[f(X_b)] = \sum\limits_{n=0}^{\infty} f^{(n)}(0)\frac{\mathbb{E}[X_b^n]}{n!} 
    \end{aligned}
    \label{eq:Ef_Taylor}
\end{equation}
and
\begin{equation}
    \begin{aligned}
     f(\mathbb{E}[X_b])=\sum\limits_{n=0}^{\infty} f^{(n)}(0)\frac{\mathbb{E}[X_b]^n}{n!} 
    \end{aligned}
    \label{eq:}
\end{equation}
where $f^{(n)}(0)$ is the $n$'th derivative of $f$ evaluated at $X_b=0$. Since $f$ is analytic everywhere, we can obtain the Jensen gap $\delta J$ by subtracting the two series term-by-term: 
\begin{equation}
    \begin{aligned}
    \delta J =\sum\limits_{n=0}^{\infty} f^{(n)}(0)\frac{\mathbb{E}[X_b]^n-\mathbb{E}[X_b^n]}{n!} 
    \end{aligned}
    \label{eq:}
\end{equation}
It is easy to see that in this series, the term corresponding to $n=0$ and $n=1$ are already zero. Since all the moments of the Binomial distribution are finite, if all the derivatives $f^{(n)}(0)$, $n=2, 3, \dots$ go to zero in the limit of infinite number of agents, then $\delta J$ will also go to zero. This is a sufficient condition for the trajectories of a deterministic map and the corresponding Binomial map to overlap in the limit of a large number of agents. It can be immediately seen that all the linear Binomial maps, that is those for which $f(X_b)=\lambda_1 X_b + \lambda_2$, satisfy the sufficient condition, and hence they are equivalent to their corresponding deterministic maps. 

For the Binomial Logistic map, 
\begin{equation}
    \begin{aligned}
    f(X_b) = \mu X_b - \mu \frac{X_b^2}{K}
    \end{aligned}
    \label{eq:}
\end{equation}
which tells us that $f^{(2)}(0)=-\mu/K$, while all other derivatives are zero. Since $f^{(2)}(0) \to 0$  as $K\to \infty$, the sufficient condition is satisfied, and the Binomial Logistic map will approach the standard Logistic map in the limit of infinite population.

For the Ricker map,
{\footnotesize
\begin{equation}
    \begin{aligned}
    f(X_b) = X_b\exp\left[\mu \left(1 - \frac{X_b}{C}\right)\right] = e^{\mu}X_b\exp\left(-\frac{\mu X_b}{C}\right)
    \end{aligned}
    \label{eq:}
\end{equation}
}
which for $n \geq 1$ gives:
{\footnotesize
\begin{equation}
    \begin{aligned}
    f^{(n)}(X_b) = \exp\left[\mu\left(1-\frac{X}{C}\right)\right]\left(-\frac{\mu}{C}\right)^{n-1}\left[n-\frac{\mu}{C}X_b\right]
    \end{aligned}
    \label{eq:}
\end{equation}
}
and hence, 
\begin{equation}
    \begin{aligned}
    f^{(n)}(0) = ne^{\mu}\left(-\frac{\mu}{C}\right)^{n-1}
    \end{aligned}
    \label{eq:}
\end{equation}
Thus, in the limit $C\to\infty$, $f^{(n)}(0)\to 0$, and hence Binomial Ricker map is also equivalent to the deterministic Ricker map in the limit of infinite population.

The sufficient condition derived for the 1-dimensional maps can be generalized to $n$-dimensional maps in a straightforward fashion. To keep the arguments simple, we will demonstrate this only for 2-dimensional discrete-time systems: 
\begin{equation}
    \begin{aligned}
    X_{t+1} &= f_1(X_{t}, Y_{t})\\
    Y_{t+1} &= f_2(X_{t}, Y_{t})
    \end{aligned}
    \label{eq:}
\end{equation}
We have not explicitly written the parameters $\Theta$ in the equations for notational simplicity. The corresponding Binomial map is given by:
\begin{equation}
    \begin{aligned}
    &\mathbb{E}[X_b(t+1)] = f_1(X_b(t), Y_b(t))\\
    &\mathbb{E}[Y_b(t+1)] = f_2(X_b(t), Y_b(t))\\
    &X_b(t+1)  = \textrm{Binomial}\left(A_{X,t}, \frac{\mathbb{E}[X_b(t+1)]}{A_{X,t}}\right)\\
    &Y_b(t+1)  = \textrm{Binomial}\left(A_{Y,t}, \frac{\mathbb{E}[Y_b(t+1)]}{A_{Y,t}}\right)\\
    \end{aligned}
    \label{eq:type1_2d_bds}
\end{equation}
Suppose we start both these systems from the same initial state, then we want to understand how much they will differ in future, if at all. Since the second system is stochastic, the better question is: if we start both the systems in the same initial state $(X_0, Y_0)$ at $t=0$, how much the deterministic system's output at later times will differ from the \emph{average} of the stochastic trajectories generated by the second system. Since we have started both the systems from the same initial condition, from the second system's equations, we see that:
\begin{equation}
    \begin{aligned}
    X(1) = \mathbb{E}(X_b(1))\quad \textrm{and}\quad Y(1) = \mathbb{E}(Y_b(1))
    \end{aligned}
    \label{eq:}
\end{equation}
Thus, the average of the stochastic trajectories at time $t=1$ is same as the value of the deterministic system. However, at $t=2$, the output of the deterministic map is: 
\begin{equation}
    \begin{aligned}
X(2) = f_1(X(1), Y(1))&=f_1(\mathbb{E}[X_b(1)], \mathbb{E}[Y_b(1)])\\
Y(2) = f_2(X(1), Y(1))&=f_2(\mathbb{E}[X_b(1)], \mathbb{E}[Y_b(1)])\\
    \end{aligned}
    \label{eq:DDS_expectation}
\end{equation}

On the other hand, the average of the stochastic trajectories is given by:
\begin{equation}
    \begin{aligned}
    \mathbb{E}[X_b(2)] &= \mathbb{E}[\mathbb{E}[X_b(2) | X_b(1), Y_b(1)]] =  \mathbb{E}[f_1(X_b(1), Y_b(1))]\\
    \mathbb{E}[Y_b(2)] &= \mathbb{E}[\mathbb{E}[Y_b(2) | X_b(1), Y_b(1)]] =  \mathbb{E}[f_2(X_b(1), Y_b(1))]\\
    \end{aligned}
    \label{eq:BDS_expectation}
\end{equation}
where we have used the \emph{Law of total expectation}.
Since in general $f(\mathbb{E}[X], \mathbb{E}[Y]) \neq \mathbb{E}(f(X, Y))$, we see from Eqs(\ref{eq:DDS_expectation}) and (\ref{eq:BDS_expectation}) that $\mathbb{E}[X_b(2)]\neq X_2$ and $\mathbb{E}[Y_b(2)]\neq Y_2$. 

To derive the sufficient condition, similar to the 1d case, we will assume that both $f_1$ and $f_2$ are analytic. The Jensen gaps for $X$ and $Y$ can then be written as:
\begin{widetext}
\begin{equation}
    \begin{aligned}
    \delta J_X &= \sum\limits_{n=0}^{\infty} \frac{1}{n!}\left[\left\{\mathbb{E}[X_b]\frac{\partial}{\partial X} + \mathbb{E}[Y_b]\frac{\partial}{\partial Y} \right\}^nf_1(0, 0) \quad - \mathbb{E}\left[\left\{X_b\frac{\partial}{\partial X} + Y_b\frac{\partial}{\partial Y} \right\}^nf_1(0, 0)\right]\right]\\
    \delta J_Y &= \sum\limits_{n=0}^{\infty} \frac{1}{n!}\left[\left\{\mathbb{E}[X_b]\frac{\partial}{\partial X} + \mathbb{E}[Y_b]\frac{\partial}{\partial Y} \right\}^nf_2(0, 0) - \mathbb{E}\left[\left\{X_b\frac{\partial}{\partial X} + Y_b\frac{\partial}{\partial Y} \right\}^nf_2(0, 0)\right]\right]\\
    \end{aligned}
    \label{eq:multi-jensen-gaps}
\end{equation}
\end{widetext}
Thus, if all the partial derivatives of all orders go to zero in the limit of infinite agents, these gaps will also go to zero. In that case, the mean of the stochastic trajectories generated by the Binomial map will be same as the deterministic trajectory of the original map. If the derivatives do not go to zero in this limit or if the functions $f_1$ and $f_2$ are not analytic at $(0, 0)$, then the Jensen gap may or may not reach to zero in the limit, and what actually happens must be analyzed for each such case separately.

\emph{Conclusion}. The formalism introduced here generalizes a class of deterministic discrete-time systems in which the variables represent counts of individuals and are therefore inherently integer-valued. By extending such systems into stochastic domain, the Binomial maps framework offers a more faithful representation of the real-world dynamics where populations are always finite. However, several foundational questions remain open, including how to appropriately define and analyze stability, sensitive dependence on initial conditions, bifurcation phenomena, and long-term behavior under stochastic fluctuations. These challenges will guide our future work as we continue to develop a comprehensive theoretical foundation for Binomial maps.

\emph{Acknowledgments}. The author would like to thank Chetan Gadgil for his insightful comments. 


\end{document}